\documentclass[thmsa,12pt]{article}

\usepackage{fancyhdr}
\usepackage{anysize}

\marginsize{1.5cm}{1.5cm}{1cm}{4cm}

\usepackage{latexsym}
\usepackage[dvips]{graphicx,color}
\usepackage{graphics}
\usepackage{pstricks}
\usepackage{epstopdf}
\usepackage[T1]{fontenc}
\usepackage{amsmath}
\usepackage{amsfonts}
\usepackage{amssymb}
\usepackage{empheq}
\usepackage[utf8]{inputenc}
\usepackage{graphicx}
\usepackage{epsfig}
\usepackage{lineno}
\usepackage{color}

\date{}
\begin{document}

\vspace{-.5cm}

\title{{\bf
Scalar field dynamics in a BTZ background with generic boundary conditions}}
\medskip

\author{Alan Garbarz$^{\dag,*}$, Joan La Madrid$^{\dag,**}$, Mauricio Leston$^{\ddag,***}$}

\maketitle

\vspace{.2cm}

\begin{minipage}{.9\textwidth}\small \it \begin{center}
    $^\dag$ Departamento de Física, FCEyN, UBA and IFIBA, CONICET,
Pabell\'on 1, Ciudad Universitaria, 1428 Buenos Aires, Argentina
 \\  \end{center}
\end{minipage}

\vspace{.2cm}

\begin{minipage}{.9\textwidth}\small \it \begin{center}
    $^\ddag$ Instituto de Astronom\'ia y F\'isica del Espacio, Pabell\'on IAFE-CONICET, Ciudad Universitaria, C.C. 67 Suc. 28, Buenos Aires
     \end{center}
\end{minipage}

\vspace{.5cm}


\begin{abstract}
We revisit the dynamics of a massive scalar field in a BTZ background taking into account the lack of global hyperbolicity of the spacetime. We approach this issue using the strategy of Ishibashi and Wald which finds a unique smooth solution as the causal evolution of initial data, each possible evolution corresponding to a positive self-adjoint extension of certain operator in a Hilbert space on the initial surface. Moreover, solutions obtained this way are the most general ones satisfying a few physically-sensible requirements. This procedure is intimately related to the choice of boundary conditions and the existence of bound states. We find that the scalar field dynamics in the (effective) mass window $-3/4\leq m_e^2\ell^2<0$ can be well-defined  within a one-parametric family of distinct boundary conditions ($-3/4$ being the conformally-coupled case), while for $m_e^2\ell^2\geq 0$ the boundary condition is unique (only one self-adjoint extension is possible). It is argued that there is no sensible evolution possible for $m_e^2\ell^2<-1$, and also shown that in the range $m_e^2\ell^2 \in [-1,-3/4)$ there is a U$(1)$ family of allowed boundary conditions, however, the positivity of the self-adjoint extensions is only motivated but not proven. We focus mainly in describing the dynamics of such evolutions given the initial data and all possible boundary conditions, and in particular we show the energy is always positive and conserved.        
\end{abstract}

\vspace{1cm}

\begin{flushleft}
\footnotesize
\parbox{\textwidth}{\mbox{}\hrulefill\\[-4pt]}
        {\scriptsize$^{*}$ E-mail: alan@df.uba.ar\\
\scriptsize$^{**}$ E-mail: joanlamadrid@gmail.com\\
\scriptsize$^{***}$ E-mail: mauricio@iafe.uba.ar}
\end{flushleft}

\newpage

\section{Introduction}

Since its discovery, the black hole solution (BTZ) to the three-dimensional Einstein's equations with a negative cosmological constant \cite{BTZ}  has been the focus of a large number of research articles. First of all, because it ``should not be there'' since every solution is locally AdS$_3$. However, black holes are, by definition, objects of a global nature, and the BTZ is a beautiful example of this fact. As reasonable solutions of general relativity, they should be considered as part of the classical phase space of solutions of gravity. Secondly, its resemblance with the four-dimensional spherically axisymmetric black holes is remarkable, notably having the same thermodynamical properties. 

Within semiclassical gravity, the phenomena of Hawking radiation \cite{Hawking} appears when fields are placed in the BTZ background, and the temperature is precisely  the surface gravity of the horizon divided by $2\pi$. In \cite{ShiraishiMaki} the stress tensor with Euclidean signature was obtained for a conformally-coupled scalar field, with Dirichlet and Neumann boundary conditions, while in \cite{Steif} and \cite{Ortiz} the stress tensor was computed by the method of images. In the last two references a conformally-coupled scalar field was studied and the crucial fact of lack of global hyperbolicity of the spacetime was stated and resolved by imposing boundary conditions on the scalar field at infinity. In \cite{Steif} the ``transparent'' boundary condition of \cite{Isham} was considered, while in \cite{Ortiz} Dirichlet and Neumann boundary conditions 
were imposed, also taking advantage of the previous results of \cite{Isham}. In \cite{Isham} the quantization of a massive or massless conformally-coupled scalar field in AdS spacetime has been considered, and the possible boundary conditions resulted from the requirement that a set of charges, constructed from the stress tensor, should be conserved. It is however not evident that this requirement guarantees a causal unique evolution of initial data, a feature that is generally desirable, and necessary to develop a canonical quantization.   

In this paper we revisit in detail the possible boundary conditions that can be imposed on the free real massive scalar field when propagating on a static BTZ background. We focus on the causal evolution of initial conditions on a spacelike surface $\Sigma$, by using the strategy first developed by Wald \cite{Wald} and continued by Ishibashi and Wald \cite{IW2,IW3}. This prescription for the evolution involves the study of (positive self-adjoint extensions of) an operator $A$ on a Hilbert space $\mathcal{H}$ on $\Sigma$. Different operators $A$ correspond with different boundary conditions). We differ the technical details for the next section, and prefer to highlight the virtues of such presription, studied in \cite{IW2}. Given a set of initial conditions $(\phi_0,\dot\phi_0) \in C_0^\infty(\Sigma)\times C_0^\infty(\Sigma)$ the prescribed evolution gives an element $\phi_t$ of $\mathcal{H}$, such that there is a unique smooth solution $\Phi$ of the equation of motion on the whole spacetime such that $\Phi|_{\Sigma_t}=\phi_t$ and $\nabla_\xi \Phi|_{\Sigma_t}=\dot\phi_t$. Here $\xi$ is the timelike Killing vector orthogonal to $\Sigma$ and $\Sigma_t$ is the image of the time-translation of $\Sigma$ ``during an interval $t$'', generated by $\xi$. Thus, we can speak of a smooth solution $\Phi$ associated to a specific initial data $(\phi_0,\dot\phi_0)$. Even more, the evolution is causal in the sense that if $K_0$ is the union of the supports of $\phi_0$ and $\dot{\phi}_0$, then the support of the solution $\Phi$ is included in the union of the causal future and causal past of $K_0$: supp$(\Phi)\subset J^+(K_0)\cup J^-(K_0)$. This fact implies that if the initial data is of compact support, then the solution $\Phi$ restricted to $\Sigma_t$ is also of compact support for a small enough $t$ \cite{IW2}. 
 
Further sensible properties are shown in \cite{IW2} to be met by the prescription of \cite{Wald}, these being first that the solution $\Phi$ respects time-translation and time-inversion. Second, there exists an energy functional $\tilde{E}$ acting on the space of solutions\footnote{To be precise, the domain of $\tilde{E}$ is the space of solutions $\mathcal{V}$ which can be written as a finite sum of other solutions with compactly supported initial data \cite{IW2}. Remarkably, $\tilde{E}$ is conserved even for this enarlged set of solutions $\mathcal{V}$.} which is positive and conserved. This energy functional agrees with an energy functional defined for globally hyperbolic spacetimes (we will come back to this at the end of this work). In addition, other technical properties are satisfied by this functional that the reader can consult in \cite{IW2}. An important result of that reference is that the prescription of $\cite{Wald}$ is actually unique if all these requirements are demanded. This means that any other way of describing an evolution that respects these physically-sensible features can be cast in the form described in \cite{Wald}.       

In \cite{IW3} the evolution in AdS spacetime and possible boundary conditions at infinity were thoroughly studied for scalar, electromagnetic and gravitational perturbations. Here we repeat their analysis for the scalar field in a static BTZ background with an arbitrary mass (which will allow to consider for example the conformally-coupled case as well as other cases in a unified setting). The main difference we need to consider with respect to the AdS case, apart from the different form of the operator $A$ to be studied, is that by working on the outter region of the black hole we may need to impose boundary conditions both at the horizon and at infinity. Everything we do could be extended, in principle, to the rotating BTZ by means of \cite{Seggev}. For a recent analysis with a particular resonance condition involving the BTZ angular velocity see \cite{Herdeiro}. 

It should be pointed out that the method here employed requires the spacetime to have a timelike Killing vector field, so we are not able to explore the dynamics inside the black hole. This is an open problem. However, is it important to stress that the dynamics outside the black hole are of most interest, since they alone imply the existence of Hawking radiation \cite{Hawking, Waldradiation}, as is rigorously shown in \cite{FH}.         

The choice of boundary conditions is important not only from a canonical quantization perspective, but also in the context of the AdS/CFT correspondence \cite{Malda}. In brief, the boundary conditions on, say, a scalar field in AdS$_3$ at infinity $(x=0)$ can be schematically written as $a x^{1-\nu}+b x^{1+\nu}$, which is interpreted as giving both the expectation value of an operator $\mathcal{O}$ of a CFT living on its boundary, as well as the source $J$ coupled with such operator. For example, for Dirchlet boundary conditions ($a=0$), $a$ serves as the source at the boundary and $b=\langle\mathcal{O}\rangle$, where $\mathcal{O}$ has scaling dimension $1+\nu$. This was first pointed out in \cite{Minces}, as far as we know, where different propagators of a scalar field were computed depending on the chosen boundary  conditions (the ratio $b/a$). Choosing different boundary conditions can be interpreted, within the AdS/CFT duality, as the insertion of a relevant perturbation in the CFT given by a double-trace operator with coupling $b/a$ \cite{Wittenboundary}. Among the observations that followed, it should be noticed that, according to the correspondence dictionary, when $b/a$ is positive the double-trace perturbation is stable. For further details see \cite{Minces2} and references therein. It seems that the analysis of the dynamics on AdS of \cite{IW3} gives further insight on the considerations just mentioned within Maldacena's conjecture. In particular \cite{IW3} gave a weaker (negative) bound on $b/a$ in order to have a sensible evolution of initial data.    
Our present work can be considered in the same manner, but for a finite-temperature setting. In particular it gives the necessary results to compute analogous quantities at finite temperature, such as the two-point function on the boundary theory.

Recently, when we were culminating this article, a preprint appeared which approaches the solutions of the equation of motion in a BTZ but from a Sturm-Liouville theory point of view, focusing mainly in the classification of the end-points and computation of Wightman two-point function \cite{Dappiaggi}.

\section{Dynamics in non-globally hyperbolic static spacetimes: the AdS case}

Now that we have motivated the advantages of the prescription of \cite{Wald,IW2} to classify all possible sensible evolutions of initial data, we give provide a review of this approach. Although it works for any static non-globally spacetime, it seems better to aim directly towards of AdS \cite{IW3}. We include this here in order to make self-cointained the present work, although the reader should see those references in order to find a more detailed account. We work only with a scalar field. Let us consider AdS$(2+n)$ in global coordinates 

\begin{equation}\label{l-metric}
ds^{2} = \frac{l^{2}}{ \sin^2(x)}  \left (-dt^{2}+dx^{2}+ \cos^2(x) d\Omega_{n}^{2}\right),
\end{equation}
where infinity is at $x=0$ and the origin at $x=\pi/2$. AdS is a static spacetime since it admits a global Killing vector field $\xi=\partial_t$ orthogonal to spacelike hypersurfaces $\Sigma_t$ defined by fixing the coordinate $t$. From now on we consider one fixed spacelike surface, say at $t=0$, and just call it $\Sigma$. The Klein-Gordon equation becomes
\begin{equation}\label{eq.KG}
(\square - m_e^2) \phi =0 \Rightarrow \frac{\partial^{2}\phi}{\partial t^{2}}=-A\phi,
\end{equation}
where $A$ denotes the spacial part of the Klein-Gordon operator. $m_e$ is an effective mass which may contain a non-minmally coupling $\Xi$ to the (constant) curvature: $m_e^2=m_0^2-\Xi(n+1)(n+2)\ell^{-2}$. Ishibashi and Wald propose to consider the Hilbert space $L^2(\Sigma,||\xi||^{-1}d\Sigma)$ of square-integrable functions on $\Sigma$ with measure given by $||\xi||^{-1}d\Sigma$. The inner product coincides with the standard Klein-Gordon product for globally-hyperbolic spacetimes \cite{IW2}.

Let the initial conditions on $\Sigma$  
be $(\phi_{0}; \dot\phi_{0}) \in C_{0}^{\infty}\left(\Sigma\right) \times C_{0}^{\infty}\left(\Sigma\right)$, then the unique and globally defined solution to (\ref{eq.KG}) is
\begin{align}\label{solWald}
\phi_{t}= \cos\left(A_{E}^{1/2}t\right)\phi_{0}+A_{E}^{-1/2} \sin\left(A_{E}^{1/2}t\right)\dot{\phi_{0}},
\end{align} 
where $A_{E}$ is a positive self-adjoint extention of $A$. The domain of $A$ is taken to be $C_0^{\infty}\left(\Sigma\right)$ and so it is a symmetric operator\footnote{Actually, since the operators $\cos(A_E^{1/2})$ and $A_E^{-1/2}\sin(A_E^{1/2})$ are bounded, the initial data can belong to $L^2(\Sigma,||\xi||^{-1}d\Sigma)$. However for initial data of compact support the corresponding evolution meets the nice properties described in \cite{IW2} and succinctly in the Introduction.}. The key point is then to find these $A_E$ operators. Note that there is always at least one self-adjoint extension if $A$ is positive, the Friederich extension. For example, $A$ is positive when $m_e^2\geq 0$. 

\subsection{Positive self-adjoint extensions of $A$}

Here we summarize and give a few details about the analysis of existence of positive self-adjoint extensions of \cite{IW3} for the scalar field with arbitrary mass $m_e$ (remember that this mass can be interpreted as an effective mass containing a non-minimal coupling to the curvature). It will serve to show how the procedure is performed and also to compare later on with our results regarding the case of spinless BTZ black holes. We start by describig how self-adjoint extensions are found and then which of these are positive.

We start by using the following decomposition
\begin{equation}\label{separationofvariables}
\phi(t,r,\theta_{i}) = r^{-n/2} \sum_{m,\bar{k}} C_{m,\bar{k}} e^{i \omega t} \varphi_{m} (r) S_{m,\bar{k}}(\theta_{i}) ,
\end{equation}
where $S_{m,\bar{k}}$ are eigenfucntions of the SO$(n+1)$ Casimir differential operator: $J^{2} S_{m,\bar{k}} = m(m+n-1) S_{m,\bar{k}}$. It is being used the fact that the Hilbert space can be decomposed in irreducible representations of the SO$(n+1)$ rotation group, so it is convenient to work with these subspaces labeled by $m$. Then, if $\phi_1$ and $\phi_2$ both belong to an $m-$subspace,
\begin{equation}
(\phi_1,\phi_2)_{L^2(\Sigma,||\xi||^{-1}d\Sigma)}=\int_\Sigma \phi_1^* \phi_2 ||\xi||^{-1}d\Sigma=\int_0^{\frac{\pi}{2}} \varphi_{1m}^*(x) \varphi_{2m}(x) dx 
\end{equation}
This means we can just consider $L^2([0,\pi/2],dx)$ in the following. 

The differential equation for $\varphi_{m}$ becomes
\begin{equation}\label{EOM-Wald}
A \varphi_m=-\frac{d^{2} \varphi_{m}}{dx^{2}} + \frac{\varphi_{m}}{ \sin^{2}(x)} \left( \nu^2 - \frac{1}{4} \right) + \frac{\varphi_{m}}{ \cos^{2}(x)} \left( \sigma^2 - \frac{1}{4}\right) = \omega^{2} \varphi_{m},
\end{equation}
where 
\begin{equation}
\nu^2-\frac{1}{4}=\frac{n(n+2)}{4}+m_e^2\ell^2,\qquad \sigma^2-\frac{1}{4}=m(m+n-1)+\frac{n(n-2)}{4}
\end{equation}
For later purposes we will choose to use sometimes the following parameters that simplify notation a little bit\footnote{Here $a(b)$ equals $\zeta^\omega_{\nu,\sigma}$ for positive (negative) $\omega$ in \cite{IW3}. Also $\nu$ denotes the positive quare root of $\nu^2$} and come from the standard notation for the Gaussian hypergeometric function
\begin{equation}
a = \frac{1+\nu+\sigma+ \omega}{2}, \qquad b = \frac{1+\nu+\sigma - \omega}{2}, \qquad
c = 1+\sigma
\label{abcAdS}
\end{equation}
Notice that $c$ and the combination $a+b$ do not depend on $\omega$. 
 
The domain of $A$ is $D(A)=C^\infty_0[0,\pi/2]$, and $A$ is then a symmetric operator.  Moreover, $A$ is positive if and only if $\nu^2\geq 0$ (Proposition 3.1, \cite{IW3}). This is the Breitenlohner-Freedman (BF) bound \cite{BF} which translates in three dimensions to $m_e^2\ell^2\geq -1$.  Thus, in the case $\nu^2\leq 0$ there is no positive self-adjoint extension of $A$, implying it is impossible to define a sensible evolution of initial data, where sensible means precisely the requirements proposed in \cite{IW2}. We shall then not consider this case anymore.

Equation (\ref{EOM-Wald}) is precisely the eigenvalue equation for the operator $A$, as well as for the operator $A^\dag$, taking into account that its domain is larger. It can be obtained by a few integration by parts and using that the domain of $A$ is the space of smooth functions of compact support,
\begin{equation}
D(A^\dag)=\left\{ \varphi \in L^2([0,\pi/2],dx)\quad ;\quad \varphi' \in AC[0,\pi/2],\quad A^\dag\varphi \in  L^2([0,\pi/2],dx)\right\} 
\end{equation}
Here $AC$ means absolutely continuous.   

We will need the general solution to (\ref{EOM-Wald}) which can be described in terms of hypergeometric functions\footnote{To be precise, the second term should be replaced by eq. (147) in \cite{IW3} if $\sigma+1$ is a natural number. However it does not change the fact that it is not square-integrable and thus $B_2=0$ anyway. },
\begin{align}
\varphi_{m} = \sin^{\nu+\frac{1}{2}}(x) \cos^{\sigma+\frac{1}{2}}(x) \left[B_{1}\,  {}_{2}F_{1}(a,b;c;\cos^{2}(x)) + B_{2} \cos^{-2\sigma}(x)  {}_{2}F_{1}(a-c+1,b-c+1;2-c;\cos^{2}(x))\right],
\label{eq3}
\end{align}
however the function multplying $B_2$ is not square-integrable because of its behaviour near the origin $x=\pi/2$ and so $B_2=0$ must be set \cite{IW3}. The remaining solution can be written in terms of two other hypergeometric functions depending on $\sin^2(x)$, which is more suitable for analysing the boundary at infinity $x=0$ (if $\nu+1 \in \mathbb{N}$ other expressions apply, see \cite{IW3} and the following section):    
\begin{equation}
\varphi_{m}(x) = G_\nu(x)\left( \frac{\Gamma(c)\Gamma(c-a-b)}{\Gamma(c-a)\Gamma(c-b)}\sin^{2\nu}(x) f_{1} + \frac{\Gamma(c)\Gamma(a+b-c)}{\Gamma(a)\Gamma(b)} f_{2}\right)
\label{sol2}
\end{equation}
where we have denoted $f_{1} = {}_{2}F_{1}(a,b;1+a+b-c;\sin^{2}(x))$ and $f_{2} = {}_{2}F_{1}(c-a,c-b;1+c-a-b;\sin^{2}(x))$ to shorten notation, and also
\begin{equation}\label{Gnu}
G_\nu:=\sin^{-\nu + 1/2}(x)\cos^{\sigma + 1/2}(x)
\end{equation}
A function (\ref{sol2}) is a generic eigenfunction of $A^\dag$  with arbitrary eigenvalue $\omega^2\in \mathbb{C}$ (disregarding momentarily the domain of $A^\dag$). However, it is not square-integrable for $\nu\geq 1$, except for certain discrete \textit{real} values of $\omega$.

\subsubsection{Self-adjoint extensions}
Since $\sigma$ only depends on $n$  and the total angular momentum $m$, $\sigma=m+(n-1)/2$, it is enough to vary $\nu^2\geq 0$ in order to give quantitative different possibilities for the self-adjoint extensions of $A$ (notice that in three dimensions $\sigma=m$). Without loss of generality we shall consider $\nu\geq 0$. The formalism developed by von Neumann requires that we find the eigenvectors of $A^\dag$ with eigenvalues $\pm i$. It is acutally more convenient (and ultimately equivalent) to consider the eigenvalues $\omega^2=\pm 2i$. We have already seen above that these are the functions (\ref{sol2}) with $\omega$ replaced by $\omega_\pm:=1\pm i$. We shall call such eigenfunctions $\varphi_m^\pm$. Taking into account the dimension of the deficiency subspaces\footnote{Deficiency subspaces are the eigenspaces associated to the purely imaginary eigenvalues $\omega_\pm^2$ of $A^\dag$. If their dimensions are $n_\pm$, then there are self-adjoint extensions if and only if $n_+=n_-$. If $n_+=n_-=0$ there is only one self-adjoint extension. If $n_+=n_->0$ then there are infinite self-adjoint extensions, labelled by partial isometries from one eigenspace to the other \cite{RS}. The case $n_+=n_-=1$ is of particular relevance to us, and implies there is a $U(1)$ family of self-adjoint extensions}, we summarize next the results of \cite{IW3} regarding self-adjoint extensions.

\begin{paragraph}{$\nu^2\geq 1$:} There is no eigenvectors (\ref{sol2}) of $A^\dag$ with eigenvalues $\omega^2=\pm 2 i$ and then there is only one self-adjoint extension. It is actually the positive one, because $A$ is positive, called Friedrichs extension.
\end{paragraph}
\begin{paragraph}{$0\leq \nu^2<1$:} There is one eigenvector (\ref{sol2}) of $A^\dag$ for each $\omega^2=\pm 2i$ and so there is a $U(1)$ family of self-adjoint extensions, parameterized by a phase $e^{i\theta}$.  
\end{paragraph}
  
Physically what this means is that depending on the value of $\nu^2$ we can have the freedom of considering different boundary conditions, each giving a distinct causal evolution of initial data. Let us explain this in more detail. Let us consider the latter case above so there is one eigenvector $\varphi_m^\pm$ generating each deficiency subspace and we have a family $A_\theta$ of self-adjoint extensions (we will make explicit $\varphi_m^\pm$ shortly). The way to see each of these operators is in one-to-one relation to different boundary conditions, and that we can indeed explicitly state them, can be seen as follows. First we have to mention that if the initial data is of compact support the solution restricted to a constant $t$ surface will always be in the domain of $A_\theta$ \cite{IW2}. The domain of such self-adjoint extension is \cite{RS}  
\begin{equation}\label{domainAdS}
D(A_\theta)=D(\bar{A}) + \mathbb{C} \varphi_m^\theta
\end{equation}
where $\bar{A}$ is the closure of $A$ and $\varphi_m^\theta=\varphi_{m}^{+} + e^{i \theta} \varphi_{m}^{-}$. Since the domain of $A$ are the smooth compactly supported functions, it is then clear that the boundary values of functions in $D(A_\theta)$ are only given by the behaviour of $\varphi_m^\theta$. Since this function depends on the phase $e^{i\theta}$, this is how it is seen that the particular self-adjoint extension determines the boundary conditions. Recall what we just mentioned above that for any time $t$ the solution is in the domain of $A_\theta$ as long as the initial conditions are of compact support. For completeness let us mention that the way $A_\theta$ operates on a function in its domain is \cite{RS}
\begin{equation}
A_\theta (\varphi_0 + z \varphi_m^\theta) = \bar{A}\varphi_0 +2iz \varphi_m^+-2ize^{i\theta}\varphi_m^-,\qquad \varphi_0\in D(\bar{A}),\quad z\in\mathbb{C}.  
\end{equation}  

In order to show the explicit way the solution behaves near infinity ($x=0$), from the discussion above it is clear that we need to understand the behavior of $\varphi_m^\theta$ in that limit. First let us write

\begin{align}\label{phitheta3}
\varphi_{m}^{\theta} & = G_{\nu} \left\{ \sin(x)^{2\nu} \left[ \frac{\Gamma(c)\Gamma(c-a-b)}{\Gamma(c-a_{+})\Gamma(c-b_{+})} f_{1}^{+} + e^{i \theta} \frac{\Gamma(c)\Gamma(c-a-b)}{\Gamma(c-a_{-})\Gamma(c-b_{-})} f_{1}^{-} \right] \right. \nonumber\\
\nonumber\\
& + \left. \left[ \frac{\Gamma(c)\Gamma(a+b-c)}{\Gamma(a_{+})\Gamma(b_{+})} f_{2}^{+} + e^{i \theta} \frac{\Gamma(c)\Gamma(a+b-c)}{\Gamma(a_{-})\Gamma(b_{-})} f_{2}^{-} \right] \right\}
\end{align}
where $f_1^{\pm}$ and $f_2^\pm$ are the hypergeometric functions defined after (\ref{sol2}) and evaluated at $\omega_{\pm}=1\pm i$ (the root chosen does not change anything, since just means an interchange $a\leftrightarrow b$ which leaves the hypergeometric functions invariant). The same applies for $a_{\pm}$ and $b_{\pm}$, and recall that $c$ and $a+b$ do not depend on $\omega$, that is why in those cases we do not indicate the sign $\pm$.

Since for $\nu^2\geq  1$ there is only one positive self-adjoint extension, the remaining case to study near $x=0$  is $0<\nu<1$ ($\nu=0$ is similarly analyzed \cite{IW3}),
\begin{align}\label{asymptotic}
\varphi_{m}^{\theta} = G_{\nu} \left\{a_{\nu} + b_{\nu} \sin(x)^{2\nu}+...\right\},\qquad x\sim 0,
\end{align}
where
\begin{align}
a_{\nu} &= - \frac{2e^{i \frac{\theta}{2}}\Gamma(1+\sigma)\Gamma(\nu)}{\left|\zeta_{\nu, \,\, \sigma}^{-(1+i)}\right|\left|\Gamma\left(\zeta_{\nu, \,\, \sigma}^{-(1+i)}\right)\right|^2} \sin\left(\frac{\theta}{2} - \theta_{+\nu}\right)
\label{anu}
\end{align}

\begin{align}
b_{\nu} &= - \frac{2e^{i \frac{\theta}{2}}\Gamma(1+\sigma)\Gamma(-\nu)}{\left|\zeta_{-\nu, \,\, \sigma}^{-(1+i)}\right|\left|\Gamma\left(\zeta_{-\nu, \,\, \sigma}^{-(1+i)}\right)\right|^2} \sin\left(\frac{\theta}{2} - \theta_{-\nu}\right) = a_{-\nu},
\label{bnu}
\end{align}

and where $\theta_{\pm \nu}\in (-\pi/2,\pi/2]$ is given by,
\begin{align}
\sin \left(\theta_{\pm \nu} \right) = \frac{\pm \nu + \sigma}{\sqrt{1+\left(\pm \nu + \sigma\right)^2}}
\label{bnu}
\end{align}
We have used the notation of \cite{IW3} where $\zeta_{\nu, \,\, \sigma}^{-(1+i)}=b_+=\zeta_{-\nu, \,\, \sigma}^{-(1+i)}+\nu$. In what follows it will prove useful to keep this notation.

It is not hard to see that possible values of $\theta \in (-\pi,\pi]$ are in bijection with the real values of the ratio
\begin{align}\label{cociente2}
\frac{b_{\nu}}{a_{\nu}} = \frac{| \zeta^{-(1+i)}_{\nu,\sigma}| |\Gamma(\zeta^{-(1+i)}_{\nu,\sigma})|^2}{| \zeta^{-(1+i)}_{-\nu,\sigma}| |\Gamma(\zeta^{-(1+i)}_{-\nu,\sigma})|^2}\frac{\Gamma(-\nu)}{\Gamma(\nu)} \frac{\sin\left(\frac{\theta}{2} - \theta_{-\nu}\right)}{\sin\left(\frac{\theta}{2} - \theta_{+\nu}\right)}
\end{align}
In other words, we have a self-adjoint extension, and thus a choice of boundary condition at infinity of AdS, for a given value of $b_\nu/a_\nu \in \mathbb{R}$. 
 
\subsubsection{Positivity: discarding the bound states}
In order to have a well-defined evolution of the initial data, one must make sure that these self-adjoint extensions are positive. This does not happen for every boundary condition. As we will perform this analysis in detail for the BTZ black hole in the following section, it seems a good idea now to review the strategy laid out in the proof of Theorem 3.2 of \cite{IW3} for the case of AdS. At the end we summarize the results.

Since the operator $A_\theta$  when restricted to the domain of compactly supported smooth functions (or better, to its closure) is positive, and since the remaining part of its domain (\ref{domainAdS}) is one-dimensional (generated by $\varphi_m^\theta$), the spectral theorem for self-adjoint operators implies that the negative spectral subspace of $A_\theta$ is at most one-dimensional. Then, if this subspace is non-empty there must be an eigenvector with negative eigenvalue. We call such eigenvector a bound state $\varphi_m^\lambda$, with eigenvalue $\omega^2=-\lambda^2$ (we take $\omega=i\lambda$ with $\lambda>0$). We want to see which $\theta$'s allow for the existence of a bound state and which do not. 

Since the bound state must belong to the domain of $A_\theta$, it must necessarily have the asymptotic behavior of $\varphi_m^\theta$ for $x=0$. Then, we should compare the asympotic of (\ref{sol2}) with $\omega=i\lambda$ and the asymptotic (\ref{asymptotic}). The former is given in an analogous way to the latter, and we define accordingly the quantities for any $\omega$
\begin{align}\label{ED}
E &= \frac{\Gamma(c)\Gamma(c-a-b)}{\Gamma(c-a)\Gamma(c-b)}=\frac{\Gamma(c)\Gamma(-\nu)}{\Gamma(\zeta^{-\omega}_{-\nu,\sigma})\Gamma(\zeta^{\omega}_{-\nu,\sigma})} \nonumber\\
\nonumber\\
D &= \frac{\Gamma(c)\Gamma(a+b-c)}{\Gamma(a)\Gamma(b)}=\frac{\Gamma(c)\Gamma(\nu)}{\Gamma(\zeta^{\omega}_{\nu,\sigma})\Gamma(\zeta^{-\omega}_{\nu,\sigma})}
\end{align}
The matching of the asymptotics gives the necessary condition (now evaluating these expressions for $\omega=i\lambda$) for $\varphi_m^\lambda$ to belong in $D(A_\theta)$,
\begin{equation}
\frac{|\Gamma(\zeta^{i\lambda}_{\nu,\sigma})|^2}{|\Gamma(\zeta^{i\lambda}_{-\nu,\sigma})|^2} = \frac{\sqrt{(\nu + \sigma)^2 + 1}}{\sqrt{(-\nu + \sigma)^2 + 1}} \frac{\left|\Gamma\left(\zeta_{\nu, \,\, \sigma}^{-(1+i)}\right)\right|^2}{\left|\Gamma\left(\zeta_{-\nu, \,\, \sigma}^{-(1+i)}\right)\right|^2} \frac{\sin\left(\frac{\theta}{2} - \theta_{-\nu}\right)}{\sin\left(\frac{\theta}{2} - \theta_{+\nu}\right)}
\end{equation} 
Now, the left-hand side is increasing with $\lambda$ and the minimum is at $\lambda=0$ \footnote{This can be seen as follows \cite{AS}: consider the property of the $\Gamma$ function $|\frac{\Gamma(x+i y)}{\Gamma(x)}|^2=\prod_{n=0}^\infty\left(1+\frac{y^2}{(x+n)^2}\right)^{-1}$. If we define $s:=\zeta^0_{-\nu,\sigma}$ and $y=i\lambda/2$, by using the property with both $x=s$ and $x=s+\nu$, we get $$|\frac{\Gamma(s+\nu+iy)}{\Gamma(s+iy)}|^2|\frac{\Gamma(s)}{\Gamma(s+\nu)}|^2= \prod_{n=0}^\infty\left(1+\frac{y^2}{(s+\nu+n)^2}\right)^{-1}\left(1+\frac{y^2}{(x+n)^2}\right)\geq 1$$. }. This implies that a necessary condition to have a bound state can be written as,
\begin{equation}\label{positivityconditionAdS}
\frac{b_\nu}{a_\nu}<-\frac{\Gamma(-\nu)}{\Gamma(\nu)} \frac{\Gamma(\zeta^0_{\nu,\sigma})^2}{\Gamma(\zeta^0_{-\nu,\sigma})^2}
\end{equation}
If this does not hold then there is no bound state and $A_\theta$ is a positive self-adjoint operator giving rise to a sensible propagation through AdS of initial data. That (\ref{positivityconditionAdS}) is sufficient for assuring the existence of a bound state is however not clear to us: it could be the case that $\varphi_m^\lambda$ has the requested asymptotics but nevertheless it is not in the domian of $A_\theta$. What is indeed certain is that if (\ref{positivityconditionAdS}) does not hold we can assure that the operator $A_\theta$ is positive and being also self-adjoint it gives a sensible propagation of the scalar field.

We can make an additional observation to the analysis of \cite{IW3}, regarding the appearance of bound states and the part they play in the Hilbert space on the initial surface. Let us say there is such a $\varphi_m^\lambda$, so this means that taking into account the time dependence a solution on AdS could be $\phi^\lambda_m= e^{-\lambda  t}\varphi_m^\lambda$, where $\varphi_m^\lambda$ does not depend on the time coordinate and is not of compact support (if it were it would belong to the domain of $A$ which is positive). The Klein-Gordon squared norm of such bound state is then 
\begin{align}
\left(\phi_m^\lambda, \phi_m^\lambda \right) = \int_{\Sigma}\! \left[(\phi_m^\lambda)^{*}\left(\nabla_{\mu}\phi_m^\lambda\right)-\left(\nabla_{\mu}(\phi_m^\lambda)^{*}\right)\phi_m^\lambda n^{\mu}\right] d\Sigma = \int_{\Sigma}\! \left[(-\lambda)(\phi_m^\lambda)^{*}\phi_m^\lambda+\lambda(\phi_m^\lambda)^{*}\phi_m^\lambda\right] \frac{1}{||\partial_{t}||} d\Sigma = 0 \nonumber
\end{align}
We see this state has zero norm. So a bound state cannot be induced on $\Sigma$  from  a solution of the Klein-Gordon equation in the whole spacetime, since such solution has zero norm and should be quotiented away. Since in the case of globally hyperbolic spacetimes the previous squared norm makes sense and one is interested in including the globally hyperbolic case in the evolution prescription, this shows that bound states should not appear. It might be the case that it is possible to modify the prescription of \cite{Wald,IW2} in order to allow first a finite-dimensional negative spectral subspace of $A_\theta$, and finally to redefine the Hilbert space by quotienting away these bound states. However, it would be expected that at least one of the nice requirements of \cite{IW2} will not be satisfied. We leave this as an open question.


\section{Dynamics in a BTZ  black hole}

The content of this section is the main original conitribution of the present work, apart from a few minor computations and comments in the previous section. We apply here the formalism developed in \cite{Wald,IW2,IW3} to the case of the exterior of a static BTZ black hole, and we obtain different possible boundary conditions where the causal evolution can be defined and then the quantization can be in principle achieved. 

The metric of a static BTZ outer region is given by,
\begin{equation}
ds^{2} = - \frac{r^{2}-r_{+}^{2}}{\ell^{2}}dt^{2}  + \frac{\ell^{2}}{r^{2}-r_{+}^{2}} dr^{2}+r^{2}d\phi^{2} \end{equation}
where $ r_{+} < r$, $-\infty< t  < + \infty$, and $0\leq \phi <2\pi$. 

By a change of coordintaes $x =  \coth^{-1} \left(\frac{r}{r_{+}}\right)$, we have that the horizon $r_+$ is at $x=\infty$ while infinity is at $x=0$. We already see a crucial difference with the AdS case: the range of the radial coordinate $x$ is now on the half-line (we could make a further change of coordinates to work on a finite range, but coordinate $x$ allows to keep the expressions similar to those of the previous section).  Let us again perform a separation of variables as in (\ref{separationofvariables}), but with a rescaled frequency $\omega\ell^2/r_+\rightarrow \omega$. By the same arguments as in the AdS case, the Hilbert space for a fixed angular momentum eigenspace is $L^2([0,\infty),dx)$, and the operator $A$ on such eigenspace is
\begin{align}
A\varphi_m=-\frac{\partial^{2} \varphi_{m}}{\partial x^{2}} + \frac{\varphi_{m}}{ \sinh^{2}\left(x\right)} \left[\frac{3}{4} + m_{0}^{2}l^{2} \right] + \frac{\varphi_{m}}{ \cosh^{2}\left(x\right)} \left[\frac{1}{4}  + m^{2}\left(\frac{l}{r_{+}}\right)^{2} \right] = \omega^{2} \varphi_{m}
\label{ecBTZ}
\end{align}
We find convenient, in order to keep as close as possible to the AdS case, to rewrite this as 
\begin{align}
A\varphi_m=-\frac{\partial^{2} \varphi_{m}}{\partial x^{2}} + \frac{\varphi_{m}}{ \sinh^{2}\left(x\right)} \left[\nu^{2} - \frac{1}{4} \right] + \frac{\varphi_{m}}{ \cosh^{2}\left(x\right)} \left[|\sigma|^2+\frac{1}{4}\right]  = \omega^{2} \varphi_{m}
\label{ecBTZ2}
\end{align}
where $\sigma$ may be now a complex number of absolute value $m\ell/r_+$ . We will keep the notation of the previous section, $\nu,\sigma,\omega$ and so on, in order to avoid clutter by writting subscripts 'BTZ', although the reader should understand that all these quantities are referring to the present case of the black hole.

At this point we can discuss the positivity of $A$. Event hough we aim to find positive self-adjoint extensions of $A$, recall from the AdS case that just from the positivity of $A$ it was possible to discard the region $\nu^2<0$. Even more, the positivity of $A$ was also useful to study the positivity of the self-adjoint extensions. Unfortunately, the procedure in \cite{IW3} to show that $A$ is positive for $\nu^2>0$ does not work in the case of the BTZ. We will content ourselves  to point out that as long as $\nu^2\geq1/4$, $A$ is positive on its domain of compactly supported smooth functions \footnote{We leave the rigorous treatment of the remaining cases for future work, although see the main text for a non-rigorous discussion}. To see this it is enough to decompose $(\varphi,A\varphi)$ as a sum of three positive integrals. Notice that this includes the conformally coupled case $\nu^2=1/4$. Although the positivity for $0\leq\nu^2<1/4$ (and non-positivity for $\nu^2<0$) remains to be proved, it is nevertheless possible to provide fairly convincing evidence: the operator $A$ consists of a kinetic term, which always gives a positive contribution, plus a potential term which is positive for $\nu^2\geq 1/4$, as we already pointed out, and for other values is positive for $x>\eta$, with $\eta$ some real positive number. The proof in \cite{IW3}, for the region $\nu^2<0$ consisits of looking at the potential where it is most negative, namely near the region $x=0$. And since in this region the AdS potential in $A$ has the same behaviour as the one for the BTZ case, then it is reasonable to expect that the same reasoning, with some adaptation, applies for the BTZ and one concludes that for $\nu^2<0$ (\ref{ecBTZ2}) is not positive. It seems harder to give compelling evidence for the range $0\leq\nu^2<1/4$, however it is a fact that the problematic region is close to $x=0$ where both operators look the same. For all this we will still consider the region $\nu^2\geq 0$ in what follows.       

The adjoint operator $ A^\dag$ is the same as $A$ as a differential operator, and its domain is computed in complete analogy to the AdS case,
\begin{equation}
D(A^\dag)=\left\{ \varphi \in L^2([0,\infty),dx)\quad ;\quad \varphi' \in AC[0,\infty),\quad A^\dag\varphi \in  L^2([0,\infty),dx)\right\} 
\end{equation}

The general solution of (\ref{ecBTZ2}) regular at $x=\infty$, as long as $a-b=i\omega$ is not an integer (this is discussed later), is
\begin{eqnarray}\label{BTZgeneralsolution}
\varphi_m &=& \sinh^{\nu+1/2}(x)	 \cosh^{\frac{1}{2} + \sigma}(x) \left[ B_1 \cosh^{-2a}(x) {}_2F_1(a,1+a-c,1+a-b; \cosh^{-2}(x))\right.\nonumber\\ 
 &+& \left. B_2 \cosh^{-2b}(x) {}_2F_1(b,1+b-c,1+b-a; \cosh^{-2}(x)) \right] 
\end{eqnarray}
where now $a$ and $b$ are defined as in (\ref{abcAdS}) but with $\omega$ replaced by $i\omega$, and $\sigma=im\ell/r_+$:
\begin{equation}
a = \frac{1+\nu+\sigma +i \omega}{2}, \qquad b = \frac{1+\nu+\sigma -i \omega}{2}, \qquad
c = 1 + \sigma
\label{abcBTZ}
\end{equation}

Now, it is important to stress that (\ref{BTZgeneralsolution}) represents a two-dimensional space of solutions of fixed $\omega^2 \in \mathbb{C}$, but in what follows we need square-integrable functions. By looking at the behaviour of the two linearly independent functions (\ref{BTZgeneralsolution}) near the horizon, $x=\infty$, it is straightforward to see that their absolute value goes as $e^{x\text{Im}(\omega)}$ for $B_2=0$ and $e^{-x\text{Im}(\omega)}$ for $B_1=0$. Thus, only one of them is square-integrable. We consider the squared root of $\omega^2$ with Im$(\omega)>0$ and set $B_1=0$ in (\ref{BTZgeneralsolution}). Choosing the other root amounts to the replacement $\omega\rightarrow -\omega$ and the interchange $a \leftrightarrow b$ in the former solution, which leaves it invariant. Thus we can consider only the case Im$(\omega)$>0, with $B_1=0$. 

This square-integrable solution, as long as $\nu+1 \notin \mathbb{Z}$, can be rewritten using the transformation identities for hypergeometric functions as \cite{AS},
\begin{align}
&\varphi_{m}= \tilde{G}_{\nu}(x) \left[\sinh^{2\nu}(x) \frac{\Gamma(b-a+1)\Gamma(c-a-b)}{\Gamma(1-a)\Gamma(c-a)} \psi_{1m} +  \frac{\Gamma(b-a+1)\Gamma(a+b-c)}{\Gamma(b)\Gamma(b-c+1)} \psi_{2m}\right],
\label{solbtz2}
\end{align}
where
\begin{equation}
\psi_{1m} = {}_{2}F_{1}(a,b;1+a+b-c;1-\cosh^{2}(x)),\qquad \psi_{2m}= {}_{2}F_{1}(c-a,c-b;1+c-a-b;1-\cosh^{2}(x))
\end{equation}
and 
\begin{equation}
\tilde{G}_\nu(x)=\sinh^{-\nu+1/2}(x)\cosh^{\sigma+1/2}(x)
\end{equation}
However, this is not square-integrable for all values of $\nu^2$. We have already studied the behavior near the horizon and now we have to do the same near infinity, $x=0$. From this it follows that (\ref{solbtz2}) is not square-integrable unless $\nu^2<1$, and as in AdS we arrive at the Breitenlohner-Freedman window \cite{BF},
\begin{equation}
0\leq \nu <1
\end{equation}
The case $\nu=0$ has to be considered separately, as in the AdS case, and we do it next. But first, let us say that it is not that surprising to arrive to the same condition for the mass in the BTZ case as in the AdS case since it comes from the behaviour of the functions near the (conformal) boundary, which both spacetimes have in common.

The cases $\nu+1 \in \mathbb{N}$, which include the BF bound $\nu=0$, can be considered in a similar way as in \cite{IW3}. The hypergeometric function in (\ref{BTZgeneralsolution}) multiplying $B_2$ has $z^{-1}=\cosh(x)^{-2}$ in its argument, and using the transformation formulas 15.3.10 and 15.3.12 of \cite{AS} it can be put as a function of $1-z^{-1}$, for the cases $\nu =0$ and $\nu=\mathbb{N}$ respectively. Now, this gives almost the same result as in \cite{IW3}, the main difference being that in that reference the solution is a function of $\sin^2(x)$ while in our case is a function of $\tanh^2(x)$, however in the limit $x=0$ both give the same and we conclude then as in \cite{IW3}: there is no square integrable solution for $\nu\geq 1$ (even if it is an integer) but for $\nu=0$ the solution is square-integrable. In brief, square-integrable solutions must have $0\leq \nu <1$.  This is valid for any $\omega \in \mathbb{C}$, with the exceptions when  $\omega$ is such that the denominators in (\ref{solbtz2}) diverge because of a negative integer in the Gamma functions. These particular cases are called degenerate \cite{AS} and we prefer not to analyze them in full detail \footnote{To be precise, the degenerate cases are defined for $c$ not an integer and at least one of the numbers $a,b,c-a,c-b$ is an integer. In our case this translates to $\omega=\pm m\frac{\ell}{r_+}\pm i 2n +i(\nu-1)$ or $\omega=\pm m\frac{\ell}{r_+}\pm i 2n -i(\nu-1)$, where $m\neq 0$.\label{footnotedegenerate} }.  Coming back to the $\nu=0$ case, this is done in detailed in \cite{IW3} and since we already justified that for $\nu=0$ the expansion near $x=0$ in the BTZ case is exactly the same as in the AdS case, we will omit the computations for $\nu=0$ and just state the results together with those for $\nu>0$. 

Now we have to repeat the analysis of square-integrability close to $x=\infty$ when $a-b=i\omega$ is an integer. When Im$(\omega)=0$ we have no square-integrable solution, since one solution of (\ref{ecBTZ2}) goes like a constant and the other is linear in $x$ \cite{AS}. For Im$(\omega)=n \in\mathbb{N}$ we will consider $n>0$, and $n<0$ will become evident from an exchange $a \leftrightarrow b$. If $a-b=n>0$, we have two independent solutions (see \cite{AS} and \cite{DLMF}),
\begin{align}\label{specialsolutions}
\varphi_{1+}&= z^{a-n/2} (1-z^{-1})^{\frac{1}{2}(\nu+1/2)} z^{-a} F(a,a-c+1;n+1;z^{-1})\sim z^{-n/2}  \\
\varphi_{2+}&=z^{a-n/2} (1-z^{-1})^{\frac{1}{2}(\nu+1/2)} z^{-a} \left[F(a,a-c+1;n+1;z^{-1})\log(z^{-1})+ \sum\limits_{r=1}^{\infty}\rho z^{-r}+ \sum\limits_{r=1}^{n}\tilde{\rho} z^{r}\right] \sim z^{n/2} 
\end{align} 
where $z=\cosh^2(x)$ and $\rho$ and $\tilde{\rho}$ mean coeficients of the sums that are unimportant. The point is that in region $x=\infty$ we see that $\varphi_{1+}$ is square-integrable while $\varphi_{2+}$ is not. For the case $n<0$ let us redefine $n$ to be positive and consider $-i\omega=b-a=n>0$, which ammounts to exchange $a\leftrightarrow b$ above, but the asymptotic behaviour is still the same:
\begin{align}
\varphi_{1-}&\sim z^{-n/2}  \\
\varphi_{2-}& \sim z^{n/2} 
\end{align} 
We see then that regardless the integer $a-b+1$, we have one possible square-integrable solution given by $\varphi_{1+}$ if $a-b=n>0$ and by $\varphi_{1-}=\varphi_{1+}|_{a\leftrightarrow b}$ if $b-a=n>0$. There remains to see their behaviour close to the boundary $x=0$. Proceeding as for the case $\omega\notin i\mathbb{Z}$, the computations are the same and as long as $\nu+1\notin\mathbb{N}>0$ we conclude $\nu<1$ in order to have a square-integrable solution. If $\nu=0$ we can use again formula 15.3.10 from \cite{AS} and see that indeed this is a square-integrable solution. For $\nu$ a positive integer we use formula 15.3.12 in \cite{AS} and see it is not square-integrable. So we have covered all cases, with the possible exceptions of the ones called degenerate (see footnote \ref{footnotedegenerate}).

\subsection{Self-adjoint extensions}

As in the AdS case, we need to find the deficiency subspaces. It will be more convenient to look for eigenstates of $A^\dag$ with eigenvalues $\pm 2i$. From the discussion above, we can summarize the results regarding the dimensions of the deficiency subspaces and thus the existence of self-adjoint extensions of $A$:

\begin{paragraph}{$\nu^2\geq 1$:} There are no eigenvectors (\ref{sol2}) of $A^\dag$ with eigenvalues $\omega^2=\pm 2 i$ and then there is only one self-adjoint extension. It is actually the positive one called Friedrichs extension.
\end{paragraph}

\begin{paragraph}{$0 \leq \nu^2<1$:} There is one eigenvector (\ref{sol2}) of $A^\dag$ for each $\omega^2=\pm 2i$ and so there is a U$(1)$ family of self-adjoint extensions parameterized by a phase $e^{i\theta}$.  
\end{paragraph}
  
These are the same conclusions as in the AdS case as for existence of self-adjoint extensions. However, the dynamics of the field, which depends on the boundary conditions, may be different iin the region $0\leq \nu<1$. We discuss the possible boundary conditions next. 

We have just seen that, as in the AdS case, the deficiency subspaces are one-dimensional, which implies that the self-adjoint extensions (and thus the boundary conditions) are parametrized by a phase $e^{i \theta}$. The boundary conditions satisfied by the scalar field coincide with the ones of $\varphi_{m}^{\theta} = \varphi_{m}^{+} + e^{i \theta} \varphi_{m}^{-}$ as explained in the previous section. The functions $\varphi_{m}^{+}$ and $\varphi_{m}^{-}$ correspond to solutions (\ref{solbtz2}) with eigenvalues $\omega_{\pm} = i\pm 1$. Note that the scalar field is necessarily exponentially suppressed when apporaching the horizon $x\rightarrow \infty$ given the analysis above in this region, so one is left to study in more detail the boundary conditions at the boundary of the BTZ.

The asymptotic behaviour of $\varphi_{m}^{\theta}$ at infinity ($x=0$) with $\nu\neq 0$  is (the $\nu=0$ case is similar, see \cite{IW3}),
\begin{align}\label{boundaryconditions}
&\varphi_{m}^{\theta} = \sinh^{\frac{1}{2} - \nu}(x) \left[ \tilde a_\nu + \tilde b_\nu \sinh^{2\nu}(x) + ...\right],
\end{align}
where we have defined the constants
\begin{align}\label{anubnu}
\tilde a_\nu &= -2\Gamma(\nu) e^{i\frac{\theta}{2}} \left|\frac{\Gamma(1-i\omega_-)}{\Gamma(b_-)\Gamma(\nu+1-a_-)}\right|\sin\left(\frac{\theta}{2}-\theta_a\right)\\
\tilde b_\nu &= -2\Gamma(-\nu) e^{i\frac{\theta}{2}} \left|\frac{\Gamma(1-i\omega_-)}{\Gamma(b_--\nu)\Gamma(1-a_-)}\right|\sin\left(\frac{\theta}{2}-\theta_b\right)=\tilde a_{-\nu}
\end{align}
where $\theta_a$ and $\theta_b$ are defined as
\begin{equation}
\sin\theta_{a}=-\frac{\text{Re}\left(\frac{\Gamma(1-i\omega_-)}{\Gamma(b_-)\Gamma(\nu+1-a_-)}\right)}{\left|\frac{\Gamma(1-i\omega_-)}{\Gamma(b_-)\Gamma(\nu+1-a_-)}\right|},\qquad \sin\theta_{b}=-\frac{\text{Re}\left(\frac{\Gamma(1-i\omega_-)}{\Gamma(b_--\nu)\Gamma(1-a_-)}\right)}{\left|\frac{\Gamma(1-i\omega_-)}{\Gamma(b_--\nu)\Gamma(1-a_-)}\right|}
\end{equation}
We have used that $(b_\pm)^*=\nu+1-a_\mp$. Again, as in the previous section, the parameters with upper index $+$ or $-$ are defined using $\omega_{\pm}$.   We follow the same reasoning of \cite{IW3} as described in the previous section and note that the values of $\theta$ are in one-to-one relation with the values of the ratio $\tilde b_\nu/\tilde a _\nu$, which can be any real number and also $\pm\infty$ (corresponding to $\tilde{a}_\nu=0$). This is a sort of Dirichlet condition at $x=0$, which is precisely Dirichlet in the conformally coupled case (see the discussion in \cite{IW3}). The generic values of $\tilde b_\nu/\tilde a _\nu$ correspond to Robin boundary conditions. These are the allowed boundary conditions at infinity that can be imposed on the scalar field in order to have a well-defined evolution, except for the fact that we have to check that the operator is positive for these boundary conditions. 

\subsection{Positivity: discarding BTZ bound states}

Finally we need to see if there is a range of the parameter $\theta$ where the self-adjoint extension $A_\theta$ is not positive. If this is the case then the corresponding boundary condition will not give an evolution with the properties described in \cite{IW2}. Recall from the discussion of positivity in the AdS case that it is important first to see the positivity of $A$ on its domain of compactly supported functions. We already discussed previously  in this section that this is the case provided $\nu^2\geq 1/4$ and gave reasons to believe that it is in fact positive for $\nu^2\geq 0$ and non-positive if $\nu^2<0$. So let us assume that this is in fact the case and proceed by noticing that since the deficiency indices are 1, the domain of $A_\theta$ is just the sum (as vector spaces) of the closure of $A$ and $\mathbb{C}\varphi_m^\theta$, the one-dimensional space generated by $\varphi_m^\theta$. Then by the same argument as for the AdS case, the operator is not positive if and only if there is a bound state on its domain (there is at most one), namely a state $\varphi^\lambda_m$ with negative energy $\omega^2=-\lambda^2$. Let us take the squared root $\omega=i\lambda$ with positive $\lambda$.  

In order to study the existence of such state we repeat the procedure as in the AdS case, namely looking for a necessary condition for its existence, which comes from demanding that it satisfies the boundary conditions (\ref{boundaryconditions}). As in (\ref{ED}) we define the coefficients that determine the asymptotics of any solution of (\ref{ecBTZ2}): 
\begin{equation}
\tilde{D}=\frac{\Gamma(\nu)\Gamma(1-i\omega)}{\Gamma(b)\Gamma(\nu+1-a)},\qquad
\tilde{E}=\frac{\Gamma(-\nu)\Gamma(1-i\omega)}{\Gamma(b-\nu)\Gamma(1-a)}
\end{equation}
We use this notation momentarily and these parameters should not be confused with the energy functional. In order for $\varphi_m^\lambda$ to have the asymptotics of $\varphi_m^\theta$, the value $\lambda$ needs to satisfy that $\tilde{E}/\tilde{D}|_{\omega=i\lambda}=\tilde{b}_\nu/\tilde{a}_\nu$. This translates into, 
\begin{equation}
\frac{\left|\Gamma(b_\lambda)\right|^2}{\left|\Gamma(b_\lambda-\nu)\right|^2}=\frac{\Gamma(\nu)}{\Gamma(-\nu)}\frac{\tilde{b}_\nu}{\tilde{a}_\nu}
\end{equation}
where $b_\lambda$ is $b$ evaluated at $\omega=i\lambda$. Similar to the strategy followed by \cite{IW3} in the AdS case, we will show that the left hand side is incresing with $\lambda$ and that it takes its minimum at $\lambda=0$. However, the identity used for AdS in the previous section is not useful here, and we instead take advantage of theorem 5.2 in \cite{IM} which readly gives that
\begin{equation}
\left|\frac{\Gamma(b_\lambda)}{\Gamma(b_\lambda-\nu)}\right|\geq \left|\frac{\Gamma(b_0)}{\Gamma(b_0-\nu)}\right|
\end{equation}  
which means that  a necessary condition to have a bound state is,
\begin{equation}\label{boundstatecondition}
\frac{\tilde{b}_\nu}{\tilde{a}_\nu}< -\left|\frac{\Gamma(-\nu)}{\Gamma(\nu)}\right|\left|\frac{\Gamma(b_0)}{\Gamma(b_0-\nu)}\right|^2,
\end{equation}
where we have used $\frac{\Gamma(-\nu)}{\Gamma(\nu)}<0$ for $\nu\in(0,1)$. If this inequality is not satisfied then we can claim the self-adjoint extension given by $\frac{\tilde{b}_\nu}{\tilde{a}_\nu}$ is positive. In other words, a sufficient condition for the self-adjoint extension to be positive is,
\begin{equation}\label{positivitycondition}
\frac{\tilde{b}_\nu}{\tilde{a}_\nu}\geq -\left|\frac{\Gamma(-\nu)}{\Gamma(\nu)}\right|\left|\frac{\Gamma(b_0)}{\Gamma(b_0-\nu)}\right|^2.
\end{equation}

\subsection{Energy conservation}

In this subsection we argue that the boundary conditions imposed at infinity of the BTZ black hole guarantee the energy conservation of the real free scalar field. This is stated somewhat differently in \cite{IW2} where no mention of the stress energy tensor is made, so we include here an explicit computation invlolving $T_{\mu\nu}$. According to the prescription proposed by Isibashi and Wald to define sensible dynamics in non-globally hyperbolic static spacetimes \cite{IW2}, the functional

\begin{align}
\widetilde{E} \left(\phi, \phi \right) &= \frac{1}{2} \int_{\Sigma}\! \dot{\phi_{0}}^2 ||\xi||^{-1} d\Sigma + \frac{1}{2} \int_{\Sigma}\! \phi_{0}A\phi_{0} ||\xi||^{-1} d\Sigma \nonumber\\ 
\nonumber\\
&= \frac{1}{2} \left(\dot{\phi_{0}}, \dot{\phi_{0}} \right)_{L^2} + \frac{1}{2} \left(\phi_{0}, A \phi_{0} \right)_{L^2}\nonumber\\
\end{align}

is conserved (is invariant under the time translation isometry applied to any solution \cite{IW2}). The inner product is the one defined on the Hilbert space $\mathcal{H}=L^{2}\left(\Sigma, ||\xi||^{-1}d\Sigma\right)$ and $\left(\phi_{0}, \dot{\phi_{0}}\right) \in  C_{0}^{\infty}\left(\Sigma\right)\times C_{0}^{\infty}\left(\Sigma\right)$ is real initial data for the Klein-Gordon equation. In the following calculations, we will show that $\widetilde{E} \left(\phi, \phi \right)$ corresponds to the energy of the solution $\phi$ generated from the initial data $\left(\phi_{0}, \dot{\phi_{0}}\right)$ (in this subsection $E$ and $\widetilde{E}$ here are not to be confused with the parameters defined in the previous subsections).

Let us consider the stress-energy tensor of a real scalar field
\begin{align}
T_{\mu \nu} = \nabla_{\mu}\phi \nabla_{\nu}\phi - \frac{1}{2} g_{\mu \nu} \left[\nabla_{\rho}\phi \nabla^{\rho}\phi  + m_{e}^2 \phi^2\right]
\end{align}
Let $\phi$  be the solution defined by the real initial data $\left(\phi_{|\Sigma} = \phi_{0}, \nabla_{\partial_{t}}\phi_{|\Sigma} =\dot{\phi_{0}}\right) \in  C_{0}^{\infty}\left(\Sigma\right)\times C_{0}^{\infty}\left(\Sigma\right)$. The energy of the field at $t=0$, $E_0$, can be calculated as the integral over the hypersurface of constant time ($t=0$), $\Sigma$, of the contraction of $T_{\mu \nu}$ with the time-like Killing vector field $\xi = \partial_t$ and the unit normal $n = \xi / ||\xi||$.

\begin{align}
E_0 &= \int_{\Sigma}\!  \left\{\frac{1}{2} \dot{\phi_{0}}^2 - \frac{1}{2}\left[g_{tt}g^{ij} \nabla_{i}\phi_{0} \nabla_{j}\phi_{0} + g_{tt}m_{e}^2 \phi_{0}^2 \right] \right\} \frac{1}{||\xi||} d\Sigma,
\end{align}
 where we used that the spacetime is static and the Latin indices correspond to spacelike coordintes. If $\sqrt{\widetilde{g}}$ is the determinant of the spacelike part of the metric $\widetilde{g}$, 
\begin{align*}
E_0 = & \int_{\Sigma}\!  \left\{\frac{1}{2} \dot{\phi_{0}}^2 + \frac{1}{2} \sqrt{-g_{tt}} \frac{1}{\sqrt{\widetilde{g}}} \left[\partial_{i} \left(\sqrt{-g_{tt}}\sqrt{\widetilde{g}}g^{ij}\phi_{0} \partial_{j}\phi_{0} \right) - \phi_{0} \partial_{i} \left(\sqrt{-g_{tt}}\sqrt{\widetilde{g}}g^{ij}\partial_{j}\phi_{0} \right) \right] - \frac{1}{2} g_{tt}m_{e}^2 \phi_{0}^2  \right\} \frac{1}{||\xi||} d\Sigma.
\end{align*}
Integrating by parts, the boundary term vanishes because $\left(\phi_{0}, \dot{\phi_{0}}\right) \in  C_{0}^{\infty}\left(\Sigma\right)\times C_{0}^{\infty}\left(\Sigma\right)$ so
\begin{align}
E_0 &= \int_{\Sigma}\!  \left\{\frac{1}{2} \dot{\phi_{0}}^2 - \frac{1}{2} \sqrt{-g_{tt}} \frac{1}{\sqrt{\widetilde{g}}} \phi_{0} \partial_{i} \left(\sqrt{-g_{tt}}\sqrt{\widetilde{g}}g^{ij}\partial_{j}\phi_{0} \right) - \frac{1}{2} g_{tt}m_{e}^2 \phi_{0}^2 \right\} \frac{1}{||\xi||}d\Sigma
\end{align}
Now, the differential operator $A$ can be expressed as $A = -||\xi|| \frac{1}{\sqrt{\widetilde{g}}}\partial_{i}\left(||\xi||\sqrt{\widetilde{g}}\hspace{0.1em} g^{ij}\partial_{j}\right) + ||\xi||^2 m_{e}^{2}$; the energy of the field at $t=0$ results

\begin{align}
E_0 &= \frac{1}{2} \int_{\Sigma}\! \left\{\dot{\phi_{0}}^2 + \phi_{0}A\phi_{0} \right\} \frac{1}{||\xi||} d\Sigma = \widetilde{E} \left(\phi, \phi \right)
\end{align}
Because the prescription of Ishibashi and Wald guarantees the conservation of the functional $\widetilde{E} \left(\phi, \phi \right)$, the energy of the field must be constant too.   

To conclude this section, we highlight that the energy is conserved for every possible positive self-adjoint extension of the operator $A$. Physically, this means that for every admissible boundary condition which can be imposed to the scalar field on the BTZ black hole, no flux of energy escapes or enters through infinity even though this spacetime is not globally hyperbolic.

\section{Conclusions}

In this work we have studied the scalar field dynamics of a real massive scalar field in the exterior of a static BTZ black hole. In order to give a detailed analysis of the evolution of initial data despite the lack of global hyperbolicity, we have taken advantage of the results of Wald \cite{Wald} and Ishibashi and Wald \cite{IW2,IW3}, which allow to determine all the boundary conditions that guarantee a physically sensible evolution.  

Despite the black hole being locally AdS, the analysis of boundary conditions needed to be repeated with caution since the horizon provides a second surface (apart from the common conformal infinity) where a specific boundary condition could be needed. For example, the operator $A$ in the case of the black hole turned out to have a different (unbounded) domain, and its form is similar though different enough from the one of AdS, making the study of positivity, in particular, non-trivial. Let us summarize what we find.   

Our results show that for $m_e^2\ell^2\geq 0$ there is only one possible boundary condition, a generalized Dirichlet boundary condition. Technically, this is given by the Friedrichs self-adjoint extension of the operator $A$ in the equation of motion, which is positive. Heuristically, the effective potential at infinity is large enough to aminish the scalar field modes in that region. What is more interesting is the BF window $-1\leq m_e^2\ell^2<0$, which includes both the conformally coupled case $m_e^2\ell^2=-3/4$ and the lower bound $m_e^2\ell^2=-1$. Within this window a U(1) family of boundary conditions, called Robin boundary conditions, can be imposed. More precisely, at least for $-3/4\leq m_e^2\ell^2<0$, we have shown that not all of them are admissible, since the evolution meets the requirements of \cite{IW2} if the extension $A_\theta$ of $A$ is not only self-adjoint but also positive. For a scalar field in the BTZ background, in order for a self-adjoint extension $A_\theta$ to be positive, negative-energy eigenstates (bound states) should not exist. A necessary condition for the existence of such bound states was given in (\ref{boundstatecondition}), so if this inequality is not met, then $A_\theta$ must be positive. We failed to give a proof that (\ref{boundstatecondition}) also holds for the whole BF window $-1\leq m_e^2\ell^2<0$, although we provided indications that this is in fact the case.

At the end we showed that one of the consequences of having a positive self-adjoint operator governing the evolution is that the typical energy functional constructed from the stress-energy tensor is positive and conserved, for any of the admissible boundary conditions. This actually holds for solutions that are not related with compactly supported initial, but are finite linear combinations of solutions labeled by initial data (see the Introduction and \cite{IW2} for further details).

All of these results are intended to give a clear understanding of the kind of solutions that one can encounter, depending not only on the boundary conditions but on the kind of initial data. For example, if the initial data is  smooth and of compact support, the solution will remain to be so for a short enough time, but will belong to the domain of $A_\theta$ at any time. We hope that this desription can be used in order to construct a rigurous quantization of a scalar field on a BTZ, where it is important to have control over the space of solutions. Strictly speaking, for every boundary condition one would get a different phase space (space of solutions), which should be endowed with a symplectic structure, and then different canonical quantizations would arise. We also hope that our results give further insight on the interpretation of double-trace perturbations on the boundary theory, according to the AdS/CFT correspondence, as we already commented in the Introduction. For instance, the lower bound (\ref{positivitycondition}), from a bulk dynamics perspective, would allow to have a small but negative double-trace perturbation coupling on the dual CFT. 

It would be also interesting to consider the influence of different Robin boundary conditions on the various notions of conserved charges, in the lines of \cite{PS} and \cite{HIM}. In those references the comparision between holographic charges and covariant phase space charges is performed, but boundary conditions on the fields are crucial in both cases (in \cite{PS} Dirichlet boundary conditions are imposed, while in \cite{HIM} the results of \cite{IW3} are taken into account). In addition, the case of dimension three is somewhat pathological, since the Weyl tensor vanishes identically and  the analysis of \cite{HIM} needs to be adapted.

\section*{Acknowledgments}

This work was supported by UBA and CONICET. We thank Claudio Dappiaggi, G. Giribet, G. P\'erez-Nadal,  and specially Guillermo Silva for many fruitful conversations. A.G. would like to thank the kind hospitality of the CCPP at New York University during completion of this work.


\begin{thebibliography}{999}


\bibitem{BTZ} M. Ba\~nados, C. Teitelboim, J. Zanelli, \textit{The Black hole in three-dimensional space-time}, Phys. Rev. Lett. \textbf{69}  (1992) 1849; hep-th/9204099.\\
M. Ba\~nados, M. Henneaux, C. Teitelboim, J. Zanelli, \textit{Geometry of the $(2+1)$ black hole }, Phys. Rev. D \textbf{48}  (1993) 1506; arXiv:gr-qc/9302012.

\bibitem{Hawking} S. W. Hawking, \textit{Particle creation by black holes}, Comm. Math. Phys.
Volume 43, Number 3 (1975) 199.

\bibitem{ShiraishiMaki}Kiyoshi Shiraishi and Takuya Maki, \textit{Quantum fluctuation of stress tensor and black holes in three dimensions} Phys. Rev. D 49, 5286 

\bibitem{Steif} A. R. Steif, \textit{The Quantum stress tensor in the three-dimensional black hole }, Phys.Rev. D49 (1994) 585;  gr-qc/9308032.

\bibitem{Ortiz} G. Lifschytz, M. Ortiz \textit{Scalar field quantization on the (2+1)-dimensional black hole background}, Phys.Rev. D49 (1994) 1929; gr-qc/9310008. 

\bibitem{Isham} S. J. Avis, C. J. Isham, and D. Storey, \textit{ Quantum field theory in anti-de Sitter space-time}, Phys. Rev. D 18, 3565.

\bibitem{Wald} R. Wald, \textit{Dynamics in nonglobally hyperbolic, static space-times}  J. Math. Phys. 21 (1980) 2802.

\bibitem{IW2} A. Ishibashi, R. M. Wald, \textit{Dynamics in Non-Globally-Hyperbolic Static Spacetimes II: General Analysis of Prescriptions for Dynamics}, Class. Quant. Grav. 20 (2003) 3815; gr-qc/0305012.

\bibitem{IW3} A. Ishibashi, R. M. Wald, \textit{Dynamics in Non-Globally-Hyperbolic Static Spacetimes III: Anti-de Sitter Spacetime}, Class. Quant. Grav. 21 (2004) 2981; hep-th/0402184. 

\bibitem{Seggev} I. Seggev, \textit{Dynamics in stationary, nonglobally hyperbolic space-times}, Class.Quant.Grav. 21 (2004) 2651; gr-qc/0310016. 

\bibitem{Herdeiro} H. R. C. Ferreira, C. A. R. Herdeiro, \textit{Stationary scalar clouds around a BTZ black hole}; 	arXiv:1707.08133. 

\bibitem{Waldradiation} R. Wald, \textit{On particle creation by black holes}, Commun.Math.Phys. 45 (1975) 9-34.


\bibitem{FH} 	K. Fredenhagen, R. Haag \textit{On the Derivation of Hawking Radiation Associated With the Formation of a Black Hole }, Commun.Math.Phys. 127 (1990) 273. 

\bibitem{Malda} J. M. Maldacena, \textit{The Large N Limit of Superconformal Field Theories and Supergravity}, Adv. Theor. Math. Phys. 2 (1998) 231; hep-th/9711200.

\bibitem{Wittenboundary} E. Witten, \textit{Multitrace operators, boundary conditions, and AdS / CFT correspondence}; hep-th/0112258.

\bibitem{Minces} P. Minces, V. O. Rivelles, \textit{Scalar field theory in the AdS / CFT correspondence revisited}, Nucl.Phys. B572 (2000) 651-669;  hep-th/9907079.

\bibitem{Minces2} P. Minces, 	\textit{Bound states in the AdS / CFT correspondence}, Phys. Rev. D70 025011; hep-th/0402161.

\bibitem{Dappiaggi} F. Bussola, C. Dappiaggi, H. R. C. Ferreira, I. Khavkine \textit{Ground state for a massive scalar field in BTZ spacetime with Robin boundary conditions}; arXiv:1708.00271. 

\bibitem{Waldlibro} R. Wald, \textit{Quantum Field Theory in Curved Spacetime and Black Hole Thermodynamics}, Chicago Lectures in Physics (1994).

\bibitem{BF} P. Breitenlohner and D.Z. Freedman, Phys. Lett. 115 B, 197 (1982). P. Breitenlohner and D.Z. Freedman, Ann. Phys. 144, 249-281 (1982).

\bibitem{RS} M. Reed, B. Simon, \textit{Fourier Analysis, Self-Adjointness (Methods of Modern Mathematical Physics), Vol. 2}, Academic Press (1975).

\bibitem{AS} M. Abramowitz, I. A. Stegun \textit{Handbook of Mathematical Functions: with Formulas, Graphs, and Mathematical Tables}, Dover Publications (1965). 

\bibitem{DLMF} \textit{NIST Digital Library of Mathematical Functions}. http://dlmf.nist.gov/, Release 1.0.15 of 2017-06-01. F. W. J. Olver, A. B. Olde Daalhuis, D. W. Lozier, B. I. Schneider, R. F. Boisvert, C. W. Clark, B. R. Miller, and B. V. Saunders, eds.

\bibitem{IM} M. E. H. Ismail and M. E. Muldoon, \textit{Inequalities and monotonicity properties
for gamma and q-gamma functions}, pp. 309-323 in R. V. M. Zahar, ed., Approximation
and Computation: A Festschrift in Honor of Walter Gautschi, ISNM,
vol. 119, Birkh\"auser, Boston-Basel-Berlin, 1994.

\bibitem{PS} I. Papadimitriou, K. Skenderis, \textit{Thermodynamics of asymptotically locally AdS spacetimes}, JHEP 0508 (2005) 004;  hep-th/0505190.

\bibitem{HIM} S. Hollands, A. Ishibashi, D. Marolf, \textit{Comparison between various notions of conserved charges in asymptotically AdS-spacetimes}, Class.Quant.Grav. 22 (2005) 2881-2920 ; hep-th/0503045.  

\end{thebibliography}
\end{document}